\documentclass[aps,prl,a4paper,twocolumn,10pt,
longbibliography,superscriptaddress]{revtex4-1}
\usepackage{amsmath}
\usepackage{amssymb}
\usepackage{mathtools}
\usepackage{graphics}
\usepackage{comment}
\usepackage{tikz}
\usepackage[normalem]{ulem}
\usepackage{multirow}

\newcommand{\beq}{\begin{equation}}
\newcommand{\eeq}{\end{equation}}
\newcommand{\bei}{\begin{itemize}}
\newcommand{\eei}{\end{itemize}}
\newcommand{\ben}{\begin{enumerate}}
\newcommand{\een}{\end{enumerate}}

\newcommand{\bk}{{\mathbf k}}
\newcommand{\bl}{{\mathbf l}}

\newcommand{\be}{{\mathbf e}}

\newcommand{\bq}{{\mathbf q}}
\newcommand{\br}{{\mathbf r}}

\usepackage{hyperref}
\hypersetup{
   pdfpagemode=UseNone, 
   pdfstartpage=1,
   pdfmenubar=true,
   pdftoolbar=true,
   colorlinks = true,
   linkcolor=blue,
   citecolor=blue,
   urlcolor=blue,
   bookmarksopen=false
 }

\definecolor{darkblue}{rgb}{0.,0.24,0.8}
\definecolor{britishracinggreen}{rgb}{0.0, 0.26, 0.15}
\definecolor{darkgreen}{rgb}{0,0.60,.2}

\def\be{\begin{equation}}
\def\ee{\end{equation}}

\begin{document}
\title{Superfluid response of bosonic fluids in composite optical potentials: angular dependence and Leggett's bounds}
\author{Daniel Pérez-Cruz}
\affiliation{Departament de F\'isica, Universitat Polit\`ecnica de Catalunya, Campus Nord B4-B5, E-08034 Barcelona, Spain}
\author{Grigori E. Astrakharchik}
\affiliation{Departament de F\'isica, Universitat Polit\`ecnica de Catalunya, Campus Nord B4-B5, E-08034 Barcelona, Spain}
\author{Pietro Massignan}
\affiliation{Departament de F\'isica, Universitat Polit\`ecnica de Catalunya, Campus Nord B4-B5, E-08034 Barcelona, Spain}
\date{\today}
	
\begin{abstract}
We study the superfluid response of a dilute bosonic fluid in the presence of two-dimensional composite potentials (such as triangular, Kagomé and quasiperiodic potentials, or superlattices), which may be obtained for example by superposing multiple laser beams. We first find a sufficient condition for the external potential to yield a fully isotropic superfluid response. Then, we derive analytical expressions for Leggett's upper and lower bounds to the superfluid fraction (valid in the perturbative regime) that allow us to find the optimal direction along which each bound should be measured. Finally, we solve the problem numerically, and we confirm our analytical findings.
\end{abstract}
\maketitle

Superfluidity is one of the most striking effects appearing in quantum many-body physics. It is found in a wide variety of systems, which encompass liquid helium, where it was first discovered \cite{Kapitza1938}, ultracold bosons \cite{Bloch2008}, strongly-interacting fermions \cite{Ketterle2008} and even very hot neutron stars \cite{Baym1969}, and it is tightly linked to superconductivity of solid-state materials. 
The extreme versatility of ultracold atom platforms made it possible to study and characterize in depth this crucial phenomenon \cite{Pethick_Smith_book_2008}.

The two-fluid model, first introduced by Gorter and Casimir in Ref.~\cite{Gorter1934} and later formalized within Landau's hydrodynamic formulation \cite{Landau1941,Tisza1938,Tisza1940_1,Tisza1940_2}, predicts that a Galilean-invariant bosonic quantum fluid in its ground state must be fully superfluid. 
When Galilean invariance is broken, however, a normal component can emerge even at zero temperature.
This phenomenon has been studied in a variety of systems, such as condensates in optical lattices \cite{Tao2023,Chauveau2023,Perez2025,Rabec2025}, where the Mott insulators are a prime example \cite{Fisher1989,Greiner2002}, supersolids \cite{Sepulveda2010,Pezze2023,Biagioni2023,Blakie2024}, spin-orbit coupled condensates \cite{Martone2012,Geier2023} or disordered systems \cite{Gaul2011,Geier2025,Perez2025,Muller2026}.

Currently, intense research is devoted to characterizing the degree of superfluidity present in a system. This information is encoded in the so-called superfluid fraction tensor. A significant breakthrough in the field occurred when Anthony Leggett proposed an upper and a lower bound for this quantity~\cite{Leggett1970,Leggett1998}. The upper bound is variational and therefore rigorous, whilst the lower bound comes from a heuristic argument and applies only to weakly-interacting systems.
Recent work has explored their applicability in bosonic~\cite{Perez2025, Chauveau2023, Blakie2024, Rabec2025, Tao2023} and fermionic systems~\cite{Orso2024}, finding Leggett's bounds to be of great use as a practical and accurate estimator of the superfluid fraction.

In this work, we demonstrate that many widely-employed composite optical potentials which feature a discrete rotational invariance (such as square, triangular, Kagomé and certain quasicrystals, or superlattices) nonetheless yield a fully-isotropic superfluid response. Leggett's bounds have instead a strong angular dependence, and here we identify the optimal directions along which the bounds are tightest. Finally, we confirm our analytical predictions by comparing them to numerical studies of relevant experimental configurations. 

{\it Superfluid fraction and perturbation theory.}---
To study the superfluid fraction of a dilute Bose-Einstein condensate (BEC) in the presence of an external static potential $V(\mathbf{r})$, we consider an additional weak potential that drags the normal component of the condensate with velocity $\mathbf{v}_0$. According to linear response theory, a weak drag couples only to the normal fraction, while the part of the system unaffected by the drag constitutes the superfluid fraction~\cite{Fisher1973}.
More specifically, given a system of $N$ particles of mass $m$, its superfluid fraction is a tensor defined through the response along the direction $\alpha$ to a perturbation moving with velocity $v_{\beta}$ along $\beta$:
\begin{equation}
\label{fs_def}
 f_{\alpha\beta}=1-\lim_{v\rightarrow 0}\langle \hat{P}_\alpha\rangle/(Nmv_{\beta}),
\end{equation}
where $\langle \hat{P}_\alpha\rangle$ is the mean total momentum along $\alpha$. 

In the frame co-moving with the dragging potential, the Gross-Pitaevskii equation (GPE) reads
\begin{equation}
\label{GP}
    \frac{(-i\hbar \nabla - m\mathbf{v}_0)^2}{2m}\psi(\mathbf{r})+ V(\mathbf{r})\psi(\mathbf{r}) + g|\psi|^2 \psi(\mathbf{r}) = \mu \psi(\mathbf{r}),
\end{equation}
where $g>0$ parametrizes the inter-particle repulsion, and $\mu$ is the chemical potential. In a uniform system of density $n_0$, we have $\mu_0=g n_0$ and a healing length $\xi = \hbar^2/\sqrt{2m\mu_0}$.
In Fourier space, the co-moving GPE reads
\begin{equation}
\sum_{\mathbf{l}}\bigg[(\varepsilon_{\mathbf{l}-\mathbf{k}_0} - \mu)\delta_{\mathbf{k},\mathbf{l}} + \frac{V(\mathbf{k}-\mathbf{l})}{\sqrt{\mathcal{V}}} + \tilde{W}(\mathbf{k},\mathbf{l}) \bigg]\psi_{\mathbf{k}_0}(\mathbf{l}) = 0,
\end{equation}
with $f(\mathbf{k})
=  \mathcal{V}^{-1/2}\int d\mathbf{r}f(\mathbf{r}) e^{-i\mathbf{k}\mathbf{r}}$  the Fourier transform, $\tilde{W}(\mathbf{k},\mathbf{l}) = (g/\mathcal{V})\sum_{\mathbf{p}}\psi_{\mathbf{k}_0}^*(\mathbf{p}\!-\!\mathbf{k})\psi_{\mathbf{k}_0}(\mathbf{p}\!-\!\mathbf{l})$  the interaction term, $\varepsilon_\mathbf{k}=\hbar^2k^2/(2m)$ and $\mathcal{V}$ the system volume.
The superfluid fraction then follows from the total momentum
\begin{equation}\label{averageP}
    \langle \hat{\mathbf{P}} \rangle =  \frac{\hbar}{2i} \int \text{d}\mathbf{r} \;   \left[ \psi_{\mathbf{k}_0}^*(\mathbf{r})\, \nabla \psi_{\mathbf{k}_0}(\mathbf{r}) -\psi_{\mathbf{k}_0}(\mathbf{r})\, \nabla \psi_{\mathbf{k}_0}^*(\mathbf{r})  \right].
\end{equation}

This equation can be solved perturbatively in both the external potential and the drag wavevector $\mathbf{k}_0= m\mathbf{v}_0/\hbar$~\cite{Gaul2011,Geier2025,Astrakharchik2013}.
To linear order in both quantities the solution is $\psi_{\mathbf{k}_0}({\bf k})=\sqrt{N/\mathcal{V}}+\delta\psi_{{\bf k}_0}({\bf k})$, with 
\begin{align}
\delta\psi_{\mathbf{k}_0}(\mathbf{k}) = -\sqrt{\frac{N}{\mathcal{V}}}\frac{V(\mathbf{k})(1-\delta_{\mathbf{k},\mathbf{0}})}{(\varepsilon_k +2\mu_0)}\left(1 +\frac{\bk \cdot \bk_0}{\epsilon_k}  \right).
\label{psi_1}
\end{align}
Substituting Eq.~\eqref{psi_1} in Eq.~\eqref{averageP} yields
\begin{equation}
    \langle \hat{\mathbf{P}} \rangle =4 \sum_{\mathbf{k}} \hat{\mathbf{k}} (\hat{\mathbf{k}} \cdot (m\mathbf{v}_0))|\delta\psi_{{\bf k}_0=0}(\mathbf{k})|^2,
\end{equation}
an expression correct up to second order in $V$ and first order in $\mathbf{k}_0$. Here \(\hat{\mathbf{k}} = \mathbf{k}/k\) denotes a unit vector with components $\hat{k}_i$. 
Thus, the superfluid fraction tensor~\eqref{fs_def} can be perturbatively expressed as~\cite{Gaul2011,Geier2025,Astrakharchik2013}
\begin{equation}
\label{fs_pert}
    f_{ij} = \delta_{ij} - \frac{4}{\mathcal{V}} \sum_{\mathbf{k}}\frac{ |V(\mathbf{k})|^2 }{(\varepsilon_k + 2\mu_0)^2} \, \hat{k}_i \hat{k}_j,
\end{equation}
a result that can also be derived within the formalism of Bogoliubov theory \cite{Giorgini1994}.

\begin{table}[t]
\caption{\label{tab:lattice_configs} 
Examples of potentials formed by $M$ shells with $N_l$ travelling waves per shell. Here, $q_l$ and $V_l$ denote the wavevector and amplitude of peaks in the $l$-th shell, and $\varphi$ is the golden ratio.}
\begin{ruledtabular}
\begin{tabular}{l c c c c}
Structure & $M$ & $N_l$ & Parameters & Ref. \\
\hline
Square & 1 & 4 & $V_1$ & -- \\
Superlattice & 2 & 4 & ${V_1,V_2}$ & -- \\
Triangular/Hexagonal & 1 & 6 & $V_1$ & -- \\
Kagomé & 2 & \{6,6\} & $V_2 = V_1/2$, $q_{2} = 2q_1$ & \cite{Jo2012} \\
Five-fold quasicrystal& 2 & \{10,10\} & $V_2 = V_1$, 
$q_2 = \varphi q_1$
& \cite{Sanchez2005} \\ 
\end{tabular}
\end{ruledtabular}
\label{Table1}
\end{table}

\begin{figure}
    \centering
    \begin{minipage}[t]{0.48\linewidth}
        \includegraphics[width=\linewidth]{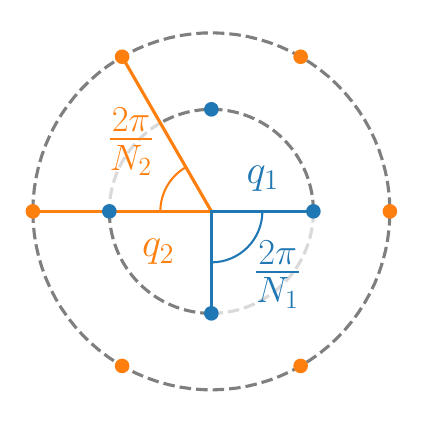}
    \end{minipage}\hfill %
    \begin{minipage}[t]{0.48\linewidth}
        \includegraphics[width=\linewidth]{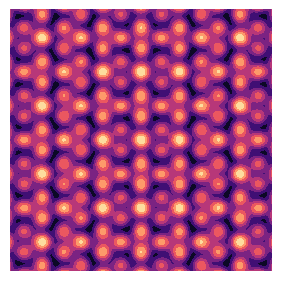}   
    \end{minipage}
    \caption{\textbf{Composite optical potentials.} Left: Typical momentum structure. The example shows a case with two shells ($M=2$), with the inner and outer shells generating square and triangular/hexagonal lattices, respectively  ($N_1=4,\, N_2=6$). Right: Resulting potential in real space.} 
    \label{fig:sketch}
\end{figure}

{\it Superfluid fraction in composite potentials}---
Superposing multiple laser beams in regular arrangements permits the generation of a large variety of exotic optical potentials. In the following, we consider the general family of two-dimensional (2D) potentials which may be expressed as
\begin{equation}
\label{family_pot}
    V(\mathbf{r}) = \sum_{l=1}^{M}\frac{V_{l}}{2}\sum_{j=1}^{N_{l}}   \exp\left(i\mathbf{q}_{lj}\cdot \mathbf{r}\right) = \sum_{l=1}^{M}V_{l}\sum_{j=1}^{N_{l}/2}   \cos\left(\mathbf{q}_{lj}\cdot \mathbf{r}\right)
\end{equation}
with $\mathbf{q}_{lj} = q_l \left[\cos(2\pi j/N_l), \sin(2\pi j/N_l)\right]$. Here $V_l$ denotes the strength of the potential in each ``shell'', and all $N_l$ need to be even for the potential to be real-valued.
Their power spectrum consists of sharp peaks forming $M$ nested regular polygons, each having $N_l$ edges:
\begin{equation}
\label{Fourier_pots}
    |V(\mathbf{k})|^2 =\frac{\mathcal{V}}{4} \sum_{l=1}^M V_l^2\sum_{j = 1}^{N_l} \delta_{\mathbf{k},\mathbf{q}_{lj}}.
\end{equation}
Notably, this family of potentials covers a wide range of experimentally relevant configurations (see Table~\ref{Table1} for a few examples, and Fig.~\ref{fig:sketch} for a specific case with two shells). 
For details on how the optical potentials generated by different configurations of laser beams can be expressed in this form, see End Matter.

\begin{figure*} 
    \centering
    \includegraphics{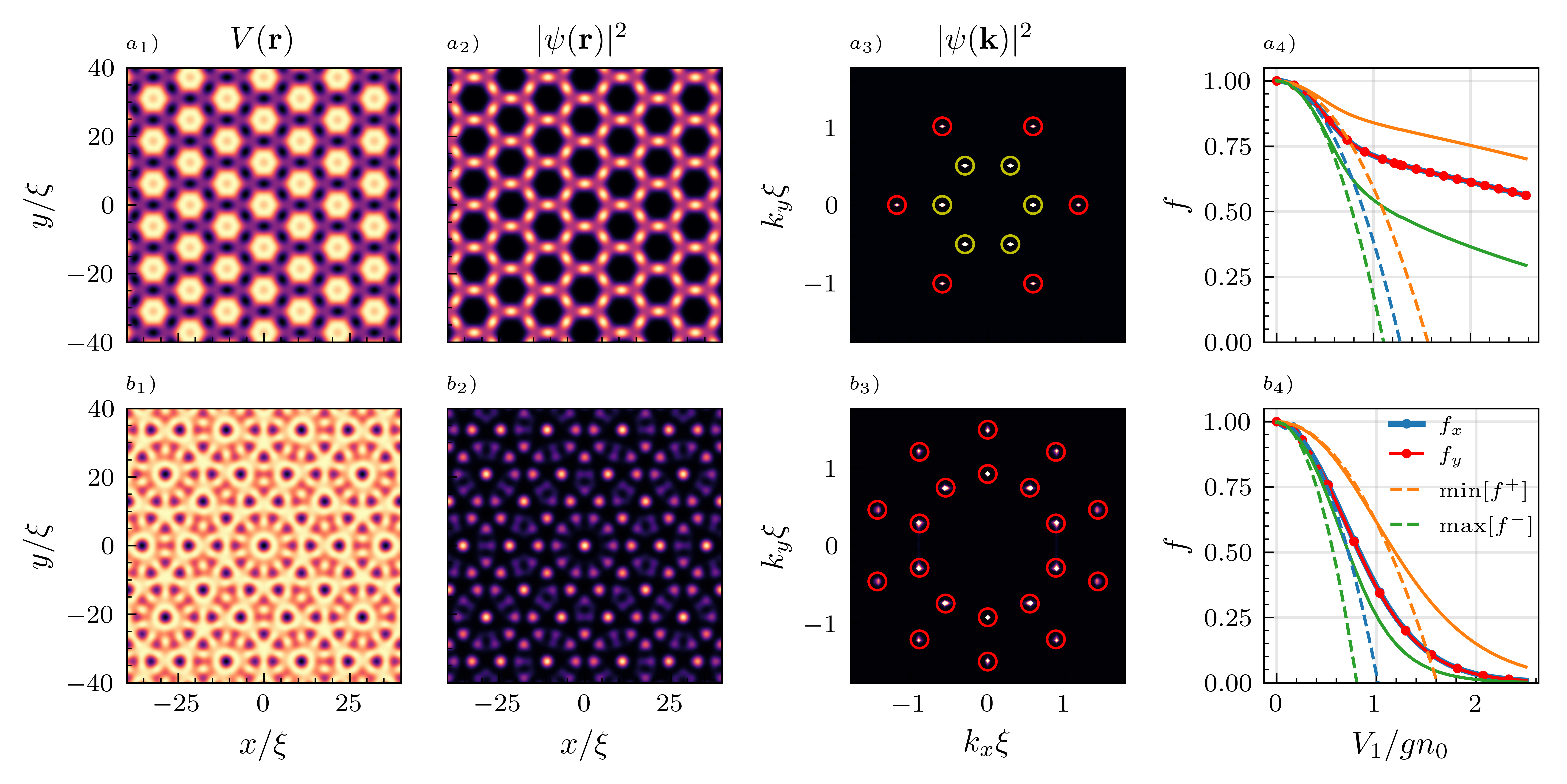} 
    \caption{\textbf{Ground state and superfluid response in composite potentials.}
Top and bottom rows correspond respectively to a BEC (modeled by the GPE, with interaction strength $gN=14400\hbar^2/m$) in a Kagomé potential and a five-fold quasicrystal (see definitions in Table~\ref{tab:lattice_configs}).
From left to right, the panels show: 
    the external potential $V(\mathbf{r})$, the ground state density, the momentum density
    (circles indicate peaks in the Fourier transform of the external potential, with yellow peaks being four times stronger than the red ones), and the superfluid fraction [dashed lines correspond to the perturbative results, Eqs.~(\ref{fs_pert},\ref{pert_upper},\ref{pert_lower})]. 
    }
    \label{fig:isotropic_fs}
\end{figure*}

In 2D the superfluid fraction is a 2*2 tensor, and the response along the direction $\hat{n} = (\cos \phi , \sin \phi)$ is 
\begin{equation}
    f(\phi) = f_{xx}\cos^2\phi + f_{yy}\sin^2\phi +(f_{xy}+f_{yx})\sin\phi \cos\phi.
\end{equation}
Since
\begin{equation}
    \frac{\partial f}{\partial \phi} = (f_{yy}-f_{xx})\sin (2\phi) +(f_{xy}+f_{yx})\cos(2\phi),
\end{equation}
$f$ becomes independent of the angle when $f_{xx} - f_{yy}=f_{xy} + f_{yx} =0$.
. 
Using the perturbative approach, for potentials of the form Eq.~\eqref{family_pot} one finds
\begin{equation}
\label{anisotropy}
\begin{aligned}
    f_{yy} - f_{xx} &= \sum_{l=1}^M \tilde{V}_{l}^2 \sum_{j=1}^{N_l} \left(\hat{q}_{lj,x}^2 - \hat{q}_{lj,y}^2 \right), \\
    f_{xy} =f_{yx} &= -\sum_{l=1}^M \tilde{V}_{l}^2 \sum_{j=1}^{N_l} \hat{q}_{lj,x} \hat{q}_{lj,y},
\end{aligned}
\end{equation}
where, as a shorthand, we have introduced the weighted amplitude 
$\tilde{V}_{l} = V_l/(\varepsilon_{q_l}+2\mu_0)$. 
Representing the two-dimensional unit wavevectors as complex roots of unity $z_l  = e^{2\pi i/N_l}$, it is easy to show that 
$(\hat{q}_{lj,x}^2 - \hat{q}_{lj,y}^2 )={\rm Re}[z_l^{2j}]$ and $\left(\hat{q}_{lj,x} \hat{q}_{lj,y}\right) ={\rm Im}[z_l^{2j}]$,
and those are themselves roots of unity. Therefore, both sums over $j$ identically vanish, and as a consequence
\begin{equation}
\label{zero_anisotropy}
    f_{yy} - f_{xx} = f_{xy} = f_{yx} = 0.
\end{equation}
Thus, the corresponding superfluid fraction is proportional to the identity matrix and therefore rotation-invariant. From this discussion follows that all potentials constructed as superpositions of regular polygons in Fourier space yield a fully isotropic response up to second order in the potential strength:
\begin{equation}
\label{eq:f}
    f = 1-\sum_{l=1}^M \frac{N_l}{2} \tilde{V}_{l}^2+O(\tilde{V}_{l}^3).
\end{equation}
We have also found that the third-order correction is zero unless $N_l$ is a multiple of three. 

As the potential strength increases, higher-order terms of the form $V(\mathbf{k}-\mathbf{l})V(\mathbf{q}-\mathbf{j})\cdots$ become relevant in the perturbative expansion, with each term involving nested differences of the initial lattice vectors. 
As discussed in the End Matter, such differences always yield vectors lying on the vertices of a regular polygon, ensuring that the isotropic response extends to all higher-order contributions. 
This demonstrates that all regular multishell potentials [as defined in Eq.~\eqref{family_pot}] yield a completely isotropic superfluid response, for arbitrarily large strengths $V_l$.

To test the validity of our findings, we solved the GPE for a 2D gas of bosons with repulsive interactions parametrized by $gN=14400\hbar^2/m$ (a value matching the experimental conditions of Ref.~\cite{Rabec2025}) in the presence of two multishell potentials: a Kagomé lattice and a five-fold quasicrystal. For the Kagomé lattice, we employed periodic boundary conditions, whereas for the quasicrystal configuration (which is aperiodic) we included steep walls at the edge of our square computational box.
Our results are shown in Fig.~\ref{fig:isotropic_fs}. 
The panels in the right column show that the superfluid response remains isotropic for arbitrarily strong potentials, even beyond the perturbative regime, in agreement with our theoretical analysis. 

{\it Optimal Leggett's bounds}---
Leggett's bounds have become an increasingly widespread tool that permits an accurate estimation of the superfluid fraction from the density profile without the need to perform dynamical measurements. 
Necessary conditions on the strength of the interactions, the temperature, or the geometry of the lattice for the bounds to be good estimators of $f$ have been discussed in previous works~\cite{Perez2025, Pizzino2025}. Leggett's bounds~\cite{Leggett1970,Leggett1998} read
\begin{equation}
\label{LBs}
    f^{+} = 
    \frac{1}{n_0} \left\langle \frac{1}{\left\langle n(\mathbf{r})\right\rangle_{\br_\perp}}\right\rangle^{-1}_{r_{\parallel}},
    \;\;\;\; 
    f^{-} = 
    \frac{1}{n_0} \left\langle \left\langle \frac{1}{n(\mathbf{r})} \right\rangle_{r_{\parallel}}^{-1} \right\rangle_{\br_\perp}.
\end{equation}
Here $n_0$ 
denotes the mean particle density, 
$r_{\parallel}$, is the direction which we call the ``measurement axis" (and coincides with the one of the weak dragging force), $\br_\perp$ represents all coordinates in the hyperplane perpendicular to $r_{\parallel}$, and $\langle \cdot \rangle_\alpha$ stands for a spatial average over coordinate $\alpha =\{r_{\parallel},\br_\perp\}$. Crucially, the bounds depend not only on the density $n({\bf r})$ but also on the measurement axis. 

In this Letter, we address two further central issues: we determine the optimal measurement axes for a given potential, and we identify which potential geometries yield the tightest bracketing. 
The angular dependence has been recently addressed in both triangular lattices~\cite{Rabec2025,Rabec2025_PhD} and supersolids~\cite{Zhang2019,Blakie2024}, and their results are consistent with our findings \footnote{It should be noted that Ref.~\cite{Blakie2024} used a definition of the lower bound which differs from ours (and from the one used in \cite{Rabec2025}), which effectively yields a shift of $\pi/2$ of the measurement direction of their lower bound.}.
As we will show, both questions can be analytically addressed in the perturbative regime discussed in the previous section. 

To study the rotational dependence of the bounds, we compute those while allowing the potential to rotate by an angle $\phi$ with respect to the weak dragging force.
By performing a perturbative expansion of the density and its inverse~\cite{Gaul2011}, we explicitly evaluate Leggett's bounds for the family of multi-shell potentials~\eqref{family_pot}. 
To second order in $\tilde{V}_l$, the upper bound exhibits the following angular dependence 
\begin{equation}
\label{pert_upper}
f^+(\phi) = 
\begin{cases} 
1 - 2\sum_{l=1}^M \tilde{V}_l^2, & \phi =2\pi j/N_l , \\
1 & \text{otherwise}
\end{cases}
\end{equation}
where $j\in \{ 1,\ldots,N_l\}$, while the lower bound is given by
\begin{equation}
\label{pert_lower}
f^-(\phi) = 
\begin{cases} 
1 - \sum_{l=1}^M (N_l-2) \tilde{V}_l^2, &\phi = 2\pi j/N_l+\pi/2, \\
1 - \sum_{l=1}^M N_l \tilde{V}_l^2, & \text{otherwise}.
\end{cases}
\end{equation}
The general perturbative expressions of the bounds can be found in the End Matter.

To find the tightest bracketing, the relevant quantity that we are interested in studying is the window between the upper and lower bounds when these take their optimal value. Simple algebra shows that 
\begin{equation}
\label{window_pert}
    \frac{f^+_{\rm opt}}{f^-_{\rm opt}} = 1 + \sum_{l=1}^M (N_l-4)\tilde{V}_l^2 .
\end{equation}
We immediately deduce that the tightest possible window ($f^-_{\rm opt} = f^+_{\rm opt} = f$) is obtained considering a square lattice, or superlattices obtained as sums of them (i.e., $N_l = 4$ for all $l$). In this case, the separability of the potential eliminates the cross-terms responsible for the gap between the optimal upper and lower bounds. Furthermore, in the End Matter we show that a result similar to~\eqref{window_pert} also applies to rectangular lattices. These do not possess an isotropic superfluid response, but they can still be tightly bound using Leggett's expressions. Note that for strong potentials the density is no longer separable and a window opens between the bounds \cite{Perez2025}.

\begin{figure}
    \centering
    \includegraphics[width=1.\columnwidth]{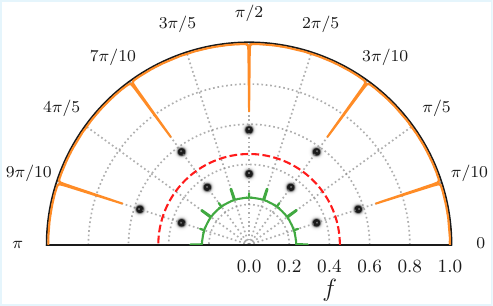}
    \caption{\textbf{Angular dependence of Leggett's bounds in a five-fold quasicrystal.}
    The plot shows the Leggett's bounds computed along different directions, for $V_1/gn \approx 0.9$ [compare with bottom-right panel of Fig.~\ref{fig:isotropic_fs}].
    The dashed red line corresponds to the numerically-extracted superfluid fraction, while the orange and green lines are its upper and lower Leggett's bounds, respectively.
    The black spots visible in the background indicate the Fourier peaks of the quasicrystal (in arbitrary units, and smoothed for visibility). 
    }
    \label{fig:bounds}
\end{figure}

Regarding the optimal axis orientation, Eqs.~(\ref{pert_upper}-\ref{pert_lower}) indicate that the optimal upper bound is attained when the dominant Fourier components of the potential are aligned with the direction of the current. On the other hand, the same analysis shows that the optimal (maximal) lower bound is found when they are disposed perpendicularly to the direction of the current. These arguments are illustrated in Fig.~\ref{fig:bounds} where the complete angular dependence of the bounds is shown. 

{\it Discussion and conclusions} ---
Back in 1894, Pierre Curie pointed out that ``the symmetries of the causes must be found in the effects'' \cite{Curie1894}. As an example, every physical property of a solid will feature all symmetries of the underlying crystalline structure. At the same time, Curie also noticed that the converse is not always true, so that the effects can be more symmetric than the causes. In this Letter we found a clear example of this, showing that the superfluid response remarkably retains full continuous rotational symmetry when the system is exposed to a large class of (arbitrarily strong) potentials which only feature a discrete invariance (such as many experimentally-relevant ones): their highly-symmetric structure in Fourier space ensures a ``geometrical protection'' to the superfluid response.

Furthermore, here we derived perturbative expressions for the Leggett's bounds applying to BECs in composite optical potentials, which give access to their full angular dependence. We showed that each of the bounds is tightest along specific directions, which are directly linked to the spectrum of the potential, and we have identified conditions where the two bounds are simultaneously tightest.

We conjecture that both the isotropy of $f$ and our findings for the angular directions giving optimal Leggett's bounds hold, in fact, for any value of the interaction strength and the temperature, as these are properties that are uniquely dictated by the geometry of the configuration.
Verifying this conjecture in the strongly interacting regime \cite{Pizzino2025} as well as in finite temperature settings \cite{Muller2026} are promising and exciting directions to be further explored.

\vspace{5mm}
\begin{acknowledgments}
We wish to thank Jean Dalibard and Jérôme Beugnon for  insightful discussions and a a careful reading of the manuscript.
We acknowledge support by the Spanish Ministerio de Ciencia, Innovación y Universidades (grants PID2023-147469NB-C21 and FPU22/03376, financed by MICIU/AEI/10.13039/501100011033 and FEDER-EU).
P.M.~further acknowledges the {\it ICREA Academia} program.
\end{acknowledgments}

\bibliography{BIBLIOGRAPHY}

\section{End Matter}
{\it Interference of travelling waves} ---
\label{app:travelling}
A potential generated by the interference of $N$ travelling waves polarized along the same direction has the expression
\begin{equation}
\label{trav_waves}
    V(\mathbf{r}) \propto \left|\sum_{j=1}^{N}e^{i\mathbf{k}_j\cdot \mathbf{r}}\right|^2.
\end{equation}
As in the main text, we will consider here only configurations whose wavevectors $\{\mathbf{k}_j\}$ are located at the edges of a regular $N$-gon centered at the origin of $k$-space. Computing the absolute value yields a sum of exponentials whose arguments are of the form $i(\mathbf{k}_i-\mathbf{k}_j)\cdot \mathbf{r}$. The Fourier transform of a potential as in Eq.~\eqref{trav_waves} is therefore composed of sharp peaks located at all possible values of $\mathbf{k}_i-\mathbf{k}_j$. Here we will show that these wavevector differences define a new set of regular polygons with even number of vertices ($N$-gons if $N$ is even, or $2N$-gons if $N$ is odd), and therefore the resulting potential may always be written as in Eq.~\eqref{family_pot}.

\begin{figure}
    \centering
    \includegraphics[width=\linewidth]{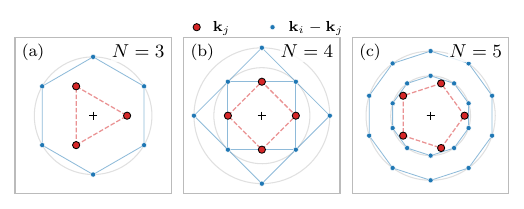}
    \caption{{\bf Spectrum of a potential generated by interfering beams} The blue dots show the location of the Fourier peaks of a potential of the form $V(\mathbf{r}) \propto \left|\sum_{j=1}^{N}e^{i\mathbf{k}_j\cdot \mathbf{r}}\right|^2$, generated by $N$ interfering beams with wavevectors ${\bf k}_j$ arranged as regular $N$-gons (red dots, with $N=3,4,5$ from left to right).}
    \label{fig:Ngons}
\end{figure}

The vertices of a regular $N$-gon (which without loss of generality we take to be of unit radius) may be expressed in complex form as $k_j = e^{2\pi i j/N}$ ($j =  1,\ldots, N$). The difference between $k_j$ and its $m^{\rm th}$-neighbor reads 
\begin{equation}
    q_{m,j} = k_{m+j}-k_{j} =  
    2 \sin \left(\frac{\pi  m}{N}\right)\exp\left[\frac{\pi i }{N}  \left(2 j+m+\frac{N}{2}\right)\right].
\end{equation}
As illustrated in Fig.~\ref{fig:Ngons}, these differences (blue dots in the sketches) form themselves $\left \lfloor{N/2}\right \rfloor$ concentric regular poligons of radii $2k\sin\left( \frac{\pi |m|}{N}\right)$, where $\left \lfloor{\ldots}\right \rfloor$ indicates the ``floor" function (or integer part). These polygons have $2N$ vertices for odd $N$, while they have only $N$ vertices when $N$ is even (because in this case one has $-q_{m,j}=q_{m,j+N/2}$).

Thus, we have shown that any potential generated by the interference of $N$ plane waves, as in Eq.~\eqref{trav_waves}, can be rewritten in the form  \eqref{family_pot}. 

As an example, consider the five-fold quasicrystal potential considered in Ref.~\cite{Sanchez2005}, which is obtained setting $N=5$ in Eq.~\eqref{trav_waves}, and corresponds to the configuration shown in the right panel of Fig.~\ref{fig:Ngons}. Following the lines of the discussion above, it is easy to see that its Fourier spectrum is composed by two concentric decagons, and the ratio of their radii is precisely the golden mean [$\sin(2\pi/5)/\sin(\pi/5)=\varphi$], in agreement with the correspoding entry in Table \ref{tab:lattice_configs}.

{\it Geometrical structure of the wavefunction spectrum} ---
In this section, we consider a potential whose power spectrum is of the form \eqref{Fourier_pots}.
We note that all $N_l$ therein are even numbers, and for simplicity we consider a potential with a single shell, and set $N=N_1$. The case with multiple shells may be treated in a very similar manner.
Naturally, the first-order correction to the wavefunction is proportional to the potential and therefore displays polygonal symmetry. Now, the second-order term is of the form
\begin{align}
    \psi_2(\bk) \propto&\sum_{\bl}\psi_1(\bk-\bl) \psi_1(\bl) \propto \sum_{\bl}V(\bk-\bl)V(\bl) \\
    \propto& \sum_{\bl} \sum_{i,j}\delta(\bk-\bl-\bq_i)\delta(\bl-\bq_j) \nonumber\\
     \propto
    &\sum_{i,j} \delta(\bk-\bq_j-\bq_i). \nonumber
\end{align}

In the first part of the End Matter we showed that terms of the form $\bq_i + \bq_j$ describe a $N$-gon.
Higher order terms will feature terms of the form $\sum_{jnl,\ldots}\delta(\bk-z_{jnl\ldots})$ with
\begin{equation}
\label{roots_comb}
    z_{jnl\ldots} = \bq_j + \bq_n +\bq_l +\ldots = \omega^j+\omega^n+\omega^l + \ldots
\end{equation}
where $\omega = e^{2\pi i /N}$ and $j,n,l$ are integers ranging from $1,\ldots, N$. The higher the order of the perturbative expansion, the more terms will appear in the sum.

To see that the resulting set has $N$-fold symmetry, we just multiply every element by $\omega$ (that in real space amounts to a rotation of $2\pi/N$) and note that we are mapping the set in itself
\begin{equation}
    \omega \{ z_{jnl\ldots}\} = \{\omega   z_{jnl\ldots}\} = \{z_{j+1,n+1,l+1,\ldots} \} = \{z_{jnl\ldots} \}.
\end{equation}
Thus, the set whose elements are of the form \eqref{roots_comb} is invariant under rotations of $2\pi/N$.

Now, we just need to show that the elements fall in concentric shells with $N$ elements. The modulus squared of the quantity \eqref{roots_comb} depends only on the difference between the indices $j,n,l,\ldots$, so if $z_{jnl}$ depends on $a$ variables, then its modulus squared depends on $a-1$ variables, and fixing these defines a shell. See for example the first part of the End Matter, where we conveniently defined the indices such that this reduction was nicely expressed. The remaining degree of freedom, which we may take to be the first index $j$, takes $N$ different values. Thus, we conclude that the set whose elements are of the form \eqref{roots_comb} describes a set of concentric $N$-gons.

{\it Perturbative expression of the bounds}  ---
\label{app:full_perturbative}
The general perturbative expression for Leggett's bounds reads
\begin{equation}
\label{pert_upper_full}
    f^+ = 1- \frac{4}{\mathcal{V}}\sum_{\mathbf{k}}\frac{|V(\mathbf{k})|^2}{(\varepsilon_k +2\mu_0)^2}\delta_{k_y,0}(1-\delta_{k_x,0})
\end{equation}
and
\begin{equation}
\label{pert_lower_full}
    f^- = 1 - \frac{4}{\mathcal{V}}\sum_{\mathbf{k}}\frac{|V(\mathbf{k})|^2}{(\varepsilon_k +2\mu_0)^2}(1-\delta_{k_x,0}).
\end{equation}
Their derivation follows the lines of that of Eq.~\eqref{fs_pert} and only depends on the first-order correction to the density function $n(\mathbf{r})$. After substituting the expression for the Fourier spectrum \eqref{Fourier_pots}, one recovers the results (\ref{pert_upper},\ref{pert_lower}). From these expressions, the window provided by the bound is found to be
\begin{equation}
\label{general_window}
     \frac{f^+(\phi)}{f^-(\phi)} =  1+\frac{4}{\mathcal{V}}\sum_{\mathbf{k}}\frac{|V(\mathbf{k};\phi)|^2}{(\varepsilon_{k}+2\mu_0)^2}(1-\delta_{k_x,0})(1-\delta_{k_y,0}).
\end{equation}
Then, for any separable potential $V(x,y) = V(x)+V(y)$, since its Fourier components lie on perpendicular axes, the bounds coincide. This result is applicable, for example, to rectangular lattices, thus generalizing result~\eqref{window_pert}.

\end{document}